\documentclass{article}
\usepackage{spconf,amsmath,graphicx,hyperref}
\usepackage{cite}
\usepackage{amssymb,amsfonts}
\usepackage{algorithmic}
\usepackage{textcomp}
\usepackage{xcolor}
\usepackage{multirow}
\usepackage{subcaption}
\usepackage{booktabs}
\usepackage{makecell}
\usepackage{multirow}
\hypersetup{breaklinks=true,hidelinks}

\title{EmphTTS: An Emphasis-Control TTS with Reinforcement Learning\\}
\name{Zirui Li \qquad Rech Silas \qquad Lauri Juvela \qquad Tom Bäckström \qquad Mikko Kurimo}
  
\address{Department of Information and Communications Engineering, Aalto University, Espoo, Finland}

\begin{document}
\ninept
\maketitle
\begin{abstract}
Generating controllable and human-like emphasis remains an open challenge in text-to-speech, even when explicit emphasis control signals are provided in the text input, limiting the communicative accuracy of synthetic speech in real-world applications. Reinforcement learning has recently shown promise for post-training TTS systems to align with human preference, yet existing methods have not been applied to word-level prosodic control. We present \textbf{EmphTTS}, a non-autoregressive TTS system that applies Group Relative Policy Optimization (GRPO) to the duration predictor with an emphasis localization reward, enabling direct optimization for word-level emphasis. 
Evaluations show that \textbf{EmphTTS} achieves the best emphasis controllability and performs the best in emphasis objective evaluation. In subjective preference tests, \textbf{EmphTTS} is significantly preferred over synthetic groundtruth and most baselines. Ablation studies show that GRPO improves emphasis realization beyond supervised-finetuning-based duration modeling and simple speaking-rate adjustment, while alleviating the mismatch between the independently trained duration predictor and TTS model.
\end{abstract}
\section{Introduction}
\label{sec:intro}

Emphasis is a prosodic feature that makes words and phrases stand out in their context, through differences in amplitude, duration, and fundamental frequency (F0)~\cite{terken2000perception}. It plays a critical role in spoken communication: the same sentence can carry entirely different meanings depending on the stressed word. For instance, \textit{``Your appointment is on Tuesday''} can single out the day (\textbf{Tuesday}, not another day), address the listener specifically (\textbf{your} appointment, not someone else's), or assert the fact against doubt (\textbf{is} on Tuesday, confirming it has not changed). In human-computer interaction, this capacity for contrastive focus is essential for dialogue repair. When a misunderstanding arises, an agent needs to be able to resolve it by emphasizing the important words, for example \textit{``Your appointment is on \textbf{Tuesday}, not Thursday''}. This fine-grained prosodic control is essential in cross-cultural settings, where automated public services and healthcare agents can reduce the cognitive load on migrants and non-native speakers. They achieve this by emphasizing key instructions and correctly stressing foreign or borrowed words~\cite{gonzalvo2014text}. Additionally, it can be used in language learning applications~\cite{Liakin19052017}, where stressing target words reinforces correct pronunciation and prosodic patterns for learners. Thus, the ability to generate appropriate word-level emphasis is essential for creating effective and trustworthy synthetic speech.

Recent TTS systems benefit from large-scale training and reinforcement learning (RL)~\cite{du2025cosyvoice,liao2026fish,hu2026qwen3}, achieving highly intelligible, expressive, and human-like speech with text-based control over speaking styles and emotions. However, recent study shows that these systems remain limited in producing correct emphasis, whether explicitly specified or inferred from context~\cite{turetzky2026knowingstress}. This suggests that high perceptual quality and broad style controllability do not necessarily transfer to fine-grained word-level prosodic control.

RL has increasingly been used to improve TTS quality and controllability. Preference-based approaches, such as Direct Preference Optimization (DPO)~\cite{zhang2024speechalign,hussain-etal-2025-koel,10888737,11460742}, optimize human-annotated preferences but require costly preference data, while automatic reward-based methods such as Group Relative Policy Optimization (GRPO)~\cite{shao2024deepseekmath} use objective metrics such as WER and speaker similarity as proxy rewards~\cite{li2026dmospeech,11462553}. Yet prior work has not targeted word-level prosodic control such as emphasis. Moreover, while GRPO fits naturally into autoregressive LM-based TTS (AR TTS), applying it to non-autoregressive (NAR) diffusion-based TTS is less direct because the sampling process is not formulated as a discrete stochastic policy. Inspired by DMOSpeech2~\cite{li2026dmospeech}, we instead exploit the probabilistic duration predictor in NAR TTS as the policy and optimize it with an emphasis-sensitive reward.


We propose \textbf{EmphTTS}, an NAR TTS system that enables word-level emphasis control by post-training its duration predictor with GRPO. We hypothesize that supervised fine-tuning (SFT) alone cannot fully model emphasis-related duration variation, while introducing a separately optimized duration predictor might lead to a mismatch with the TTS model at inference. We therefore apply GRPO to the duration predictor while keeping the TTS model fixed, directly optimizing duration decisions for emphasis realization. Experiments show that this improves objective emphasis control beyond SFT and yields strong subjective preference over most compared systems. These results highlight the potential of RL for fine-grained prosodic control in NAR TTS. Audio samples are available at \url{https://aalto-speech.github.io/EmphTTS/} and code is released at \url{https://github.com/aalto-speech/EmphTTS}.

\begin{figure*}[t]
    \centering
    \begin{subfigure}{.4\linewidth}
        \centering
        \includegraphics[width=\linewidth]{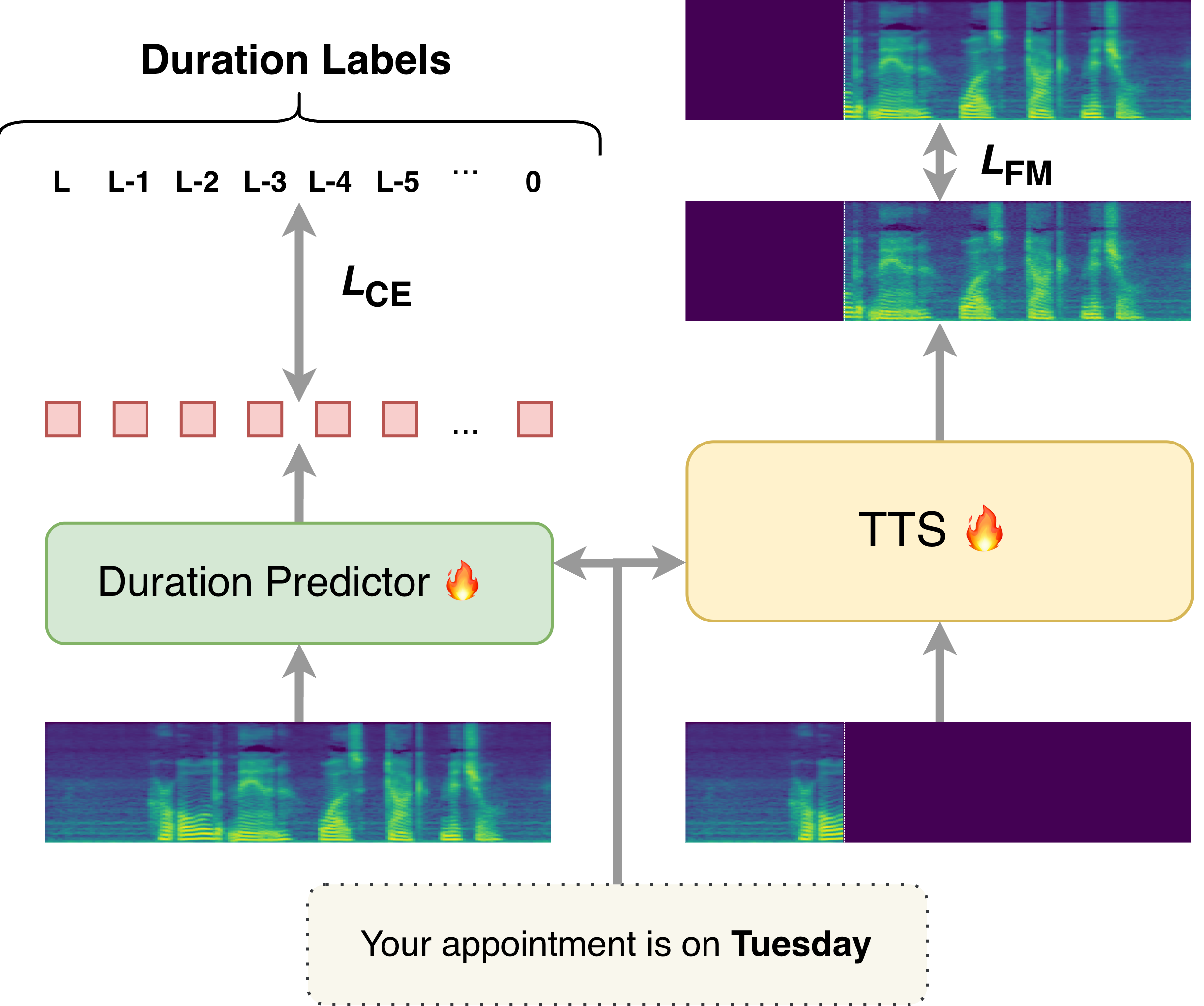}
        \caption{Supervised fine-tuning (SFT) stage.}
        \label{fig:sft-stage}
    \end{subfigure}
    \hspace{0.05\linewidth}
    \begin{subfigure}{.4\linewidth}
        \centering
        \includegraphics[width=\linewidth]{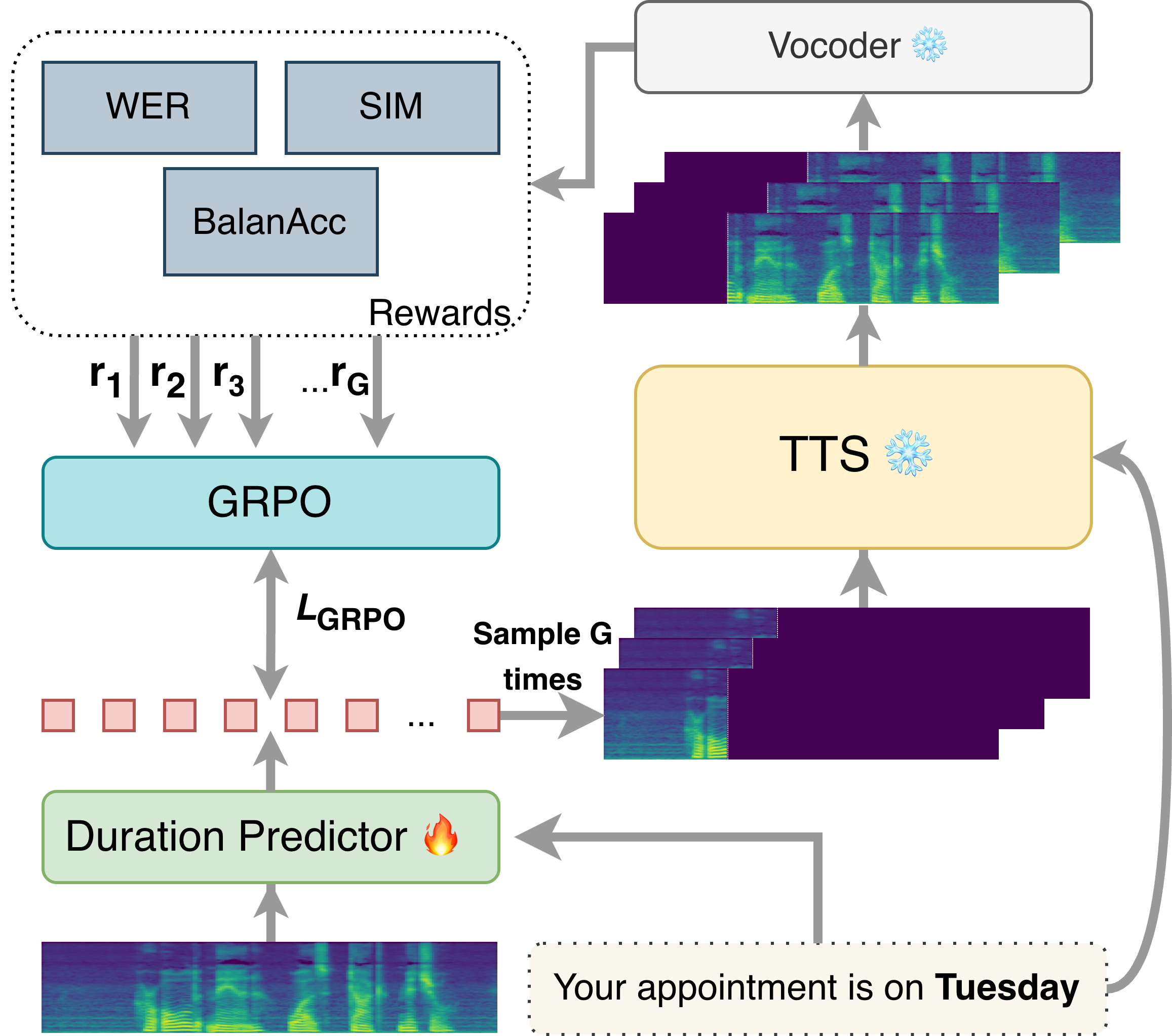}
        \caption{Group Relative Policy Optimization (GRPO) stage.}
        \label{fig:grpo-stage}
    \end{subfigure}
    \caption{The overview of the SFT (left) and GRPO (right) stage. \textbf{(a) Left}: The TTS model and the duration predictor are fine-tuned using the flow-matching objective ($L_{FM}$) and cross-entropy ($L_{CE}$). \textbf{(b) Right}: The duration predictor is trained with GRPO while the TTS model is frozen for inference.}
    \label{fig:both}
    \vspace{-5mm}
\end{figure*}

\section{Related Work}
\label{sec:relatedWork}

\subsection{Duration Modeling in TTS}

For AR TTS, duration modeling is done implicitly by predicting the \textit{end-of-sentence (EOS)} token when decoding. On the other hand, NAR TTS systems require a duration predictor to estimate the length of the speech they need to generate given the text input~\cite{kim2020glow}. Early work such as Glow-TTS~\cite{kim2020glow} and FastSpeech2~\cite{ren2021fastspeech} model duration at the phoneme level through monotonic alignment search or an external forced aligner. DiTTo-TTS~\cite{lee2025ditto} introduced a speech length predictor that is trained to predict the total generation length given the prompt speech and the target text, eliminating the hard phoneme boundaries in NAR TTS and leading to better human-likeness in the synthetic speech. Modern NAR TTS systems such as E2-TTS~\cite{eskimez2024e2} and F5-TTS~\cite{chen2025f5} eliminate the explicit duration predictor by estimating the target speech duration as
$\hat{L}_\text{target} = T_\text{target} \cdot L_\text{prompt} \,/\, T_\text{prompt}$,
where $T_\text{target}$ is the number of input text tokens, $L_\text{prompt}$ is the prompt audio length, and $T_\text{prompt}$ is the number of prompt text tokens. The ratio $L_\text{prompt} / T_\text{prompt}$ thus serves as an estimate of the average audio duration per text token in the prompt. A global speed factor $s$ can further rescale this estimate as
$\hat{L}'_\text{target}=\hat{L}_\text{target}/s$, providing coarse
utterance-level speaking-rate control. 
Since emphasis markers add only a few tokens to $T_\text{target}$, the unscaled estimate $\hat{L}_\text{target}$ often underestimates the duration required to realize emphasis-related lengthening~\cite{terken2000perception}. While global speed scaling can compensate for this at the utterance level, the appropriate scale varies with the number and locations of emphasized words, making fixed global scaling inherently coarse for word-level emphasis control.

\subsection{Reinforcement Learning in TTS}

Reinforcement Learning applied to TTS broadly follows preference-based and automatic-reward approaches. SpeechAlign~\cite{zhang2024speechalign} proposed an iterative self-improvement strategy for codec language model-based TTS using preference datasets. Koel-TTS~\cite{hussain-etal-2025-koel} applied preference alignment guided by ASR and speaker verification to encoder-decoder TTS. Emo-DPO~\cite{10888737} used DPO to improve emotional expressiveness. For diffusion-based TTS, Chen et al.~\cite{chen25b_interspeech} introduced diffusion loss-guided policy optimization to improve naturalness. While these DPO-based methods demonstrate strong alignment with human preferences, they rely on preference datasets that incur additional data-collection costs.

Group Relative Policy Optimization (GRPO)~\cite{shao2024deepseekmath} offers a complementary approach that replaces preference pairs with automatic reward signals and avoids the need for a separate value network. Liu et al.~\cite{11462553} applied GRPO to LM-based TTS using Character Error Rate (CER) and negative log-likelihood (NLL) from an ASR model as reward signals, demonstrating improvements in speech intelligibility. DMOSpeech2~\cite{li2026dmospeech} extended this to NAR TTS by applying GRPO to the duration predictor while keeping the TTS model frozen, showing substantial improvements in synthesis quality. However, both approaches optimize solely for intelligibility and speaker-level metrics: Shin et al.~\cite{11460742} showed that such transcription-centric rewards collapse prosodic variation into monotone output, and that adding speaker similarity rewards further destabilizes training. Our work avoids this by introducing an emphasis-specific reward into the GRPO objective, enabling fine-grained prosodic control and ensuring prosodic variation in the synthetic speech.

\section{Methodology}
\label{sec:method}

Our \textbf{EmphTTS} uses F5-TTS~\cite{chen2025f5} as the backbone. We first pre-train it with the standard flow-matching objective, yielding the base model $\mathcal{M}_\text{TTS}$. For duration prediction, we adopt the encoder-decoder architecture of DiTTo-TTS~\cite{lee2025ditto}. Given text $\mathbf{x}$ and prompt speech $\mathbf{y}_\text{ref}$, the duration predictor $\mathcal{M}_\text{DP}$ predicts the remaining speech length at each timestep as a categorical distribution over a maximum length $L_\text{max}$. Using the countdown target $L_t=L-t$, it is trained with
\begin{equation}
    \mathcal{L}_\text{DP}
    = -\sum_t \log p_\theta(L_t \mid \mathbf{x},\,\mathbf{y}_{<t}).
    \label{eq:dp_loss}
\end{equation}
At inference, the target duration is sampled from the distribution predicted at the end of the prompt:
\begin{equation}
    \hat{L} \sim p_\theta(L \mid \mathbf{x},\,\mathbf{y}_\text{ref}).
    \label{eq:dur_sample}
\end{equation}
Both $\mathcal{M}_\text{TTS}$ and $\mathcal{M}_\text{DP}$ are then separately fine-tuned on emphasis-labeled data, yielding $\mathcal{M}_\text{TTS-SFT}$ and $\mathcal{M}_\text{DP-SFT}$. Because the two models are optimized independently, directly combining them remains suboptimal for emphasis generation. To reduce this mismatch, we optimize $\mathcal{M}_\text{DP-SFT}$ with GRPO~\cite{shao2024deepseekmath}, while keeping $\mathcal{M}_\text{TTS-SFT}$ frozen, as shown in Fig.~\ref{fig:grpo-stage}. For each input, we sample $G$ candidate durations $\{\hat{L}_i\}_{i=1}^{G}$ and synthesize
$\hat{\mathbf{y}}_i =
\mathcal{M}_\text{TTS-SFT}
(\mathbf{x},\mathbf{y}_\text{ref},\hat{L}_i)$.
Each candidate is evaluated using WER, speaker similarity (SIM), and balanced accuracy~\cite{5597285} for emphasis localization (BalanAcc). The reward is
\begin{align}
    r(\hat{\mathbf{y}}_i) =\;
        &-\lambda_\text{WER}\,\text{WER}(\hat{\mathbf{y}}_i)
        +\lambda_\text{SIM}\,\text{SIM}(\hat{\mathbf{y}}_i,\mathbf{y}_\text{ref})
        \notag\\
        &+\lambda_\text{BalanAcc}\,\text{BalanAcc}(\hat{\mathbf{y}}_i),
    \label{eq:reward}
\end{align}
and is normalized within each group to obtain
$\hat{A}_i=(r_i-\mu_r)/\sigma_r$.

Let
$R_i(\theta)=
p_\theta(\hat{L}_i\mid\mathbf{x},\mathbf{y}_\text{ref})/
p_\text{ref}(\hat{L}_i\mid\mathbf{x},\mathbf{y}_\text{ref})$.
The clipped surrogate objective is
\begin{equation}
    h_i(\theta) =
    \min\!\left(
        R_i(\theta)\hat{A}_i,\,
        \operatorname{clip}(R_i(\theta),1-\epsilon,1+\epsilon)\hat{A}_i
    \right),
    \label{eq:grpo_clip}
\end{equation}
and the GRPO loss is
\begin{equation}
    \mathcal{L}_\text{GRPO}(\theta)
    = -\frac{1}{G}\sum_{i=1}^{G} h_i(\theta)
      + \beta D_\text{KL}(p_\theta\|p_\text{ref}),
    \label{eq:grpo}
\end{equation}
where $p_\text{ref}$ is initialized from $\mathcal{M}_\text{DP-SFT}$.

The configurations above yield six systems for evaluation. The two base systems are $\mathcal{M}_\text{TTS}$ and $\mathcal{M}_\text{TTS-DP}$, where the latter pairs $\mathcal{M}_\text{TTS}$ with $\mathcal{M}_\text{DP}$. We then consider three ablations: $\mathcal{M}_\text{TTS-SFT}$ without duration prediction, $\mathcal{M}_\text{TTS-SFT}$ with a global speed factor $s=0.9$ at inference, and $\mathcal{M}_\text{TTS-DP-SFT}$, which combines $\mathcal{M}_\text{TTS-SFT}$ with $\mathcal{M}_\text{DP-SFT}$. The speed-factor variant tests whether improvements in emphasis can be achieved by reducing the speaking rate, and we choose $s=0.9$ based on experimental observations. Finally, our proposed \textbf{EmphTTS} combines $\mathcal{M}_\text{TTS-SFT}$ with $\mathcal{M}_\text{DP-GRPO}$.

\section{Experiments}

\begin{table*}[t]
\centering
\small
\renewcommand{\arraystretch}{0.82}
\setlength{\tabcolsep}{3.5pt}
\setlength{\aboverulesep}{0.4pt}
\setlength{\belowrulesep}{0.8pt}

\caption{Objective and subjective evaluation results on \textit{Seed-TTS EN}
and \textit{TinyStress-15k}. Qwen3-TTS-VD denotes
\textit{Qwen3-TTS-Voice-Design}; $\sim$ denotes unavailable results.
Column-best results are \textbf{bold}, and the best results among our base,
ablation, and proposed systems are \underline{underlined}. 
\textbf{EmphPref L/W} reports the one-versus-rest EmphPref
comparison between \textbf{EmphTTS} and each listed system from the perspective
of \textbf{EmphTTS}: L denotes the percentage of trials in which
\textbf{EmphTTS} loses, and W denotes the percentage in which
\textbf{EmphTTS} wins. Ties are omitted and account for the remaining
percentage. $^{*}$ indicates that \textbf{EmphTTS} is significantly preferred
over the compared system under a one-sided binomial test on non-tie responses.
CMOS is reported with 95\% confidence intervals.}
\label{tab:exp}

\begin{tabular*}{\textwidth}{
@{\extracolsep{\fill}}
ll
ccc
cccc
@{}
}
\toprule

& &
\multicolumn{3}{c}{\textbf{Seed-TTS EN}} &
\multicolumn{4}{c}{\textbf{TinyStress-15k}} \\
\cmidrule(lr){3-5}
\cmidrule(lr){6-9}

& &
\textbf{WER}$\downarrow$ &
\textbf{SIM}$\uparrow$ &
\textbf{CMOS}$\uparrow$ &
\textbf{WER}$\downarrow$ &
\textbf{SIM}$\uparrow$ &
\textbf{StressLM F1}$\uparrow$ &
\shortstack{\textbf{EmphPref}\\[-1pt]\textbf{L/W (\%)}} \\
\midrule

\multicolumn{2}{c}{\textbf{Groundtruth}}
& 1.86\%
& $\sim$
& $\sim$
& \textbf{0.25\%}
& $\sim$
& 0.60
& 36/53.7$^{*}$ \\

\midrule

\multirow{4}{*}{\textbf{Baselines}}
& CosyVoice3
& 1.76\%
& 0.70
& $-0.23 \pm 0.15$
& 0.32\%
& 0.68
& 0.26
& 37.7/56.0$^{*}$ \\

& FishAudio-S2
& \textbf{0.99\%}
& 0.65
& $-0.22 \pm 0.18$
& \textbf{0.25\%}
& 0.67
& 0.41
& 33.2/56.5$^{*}$ \\

& Qwen3-TTS-VD
& 1.70\%
& $\sim$
& $\mathbf{\phantom{-}0.06 \pm 0.17}$
& 0.29\%
& $\sim$
& 0.63
& 44.8/45.7 \\

& VoxCPM2
& 1.84\%
& $\mathbf{0.75}$
& $-0.07 \pm 0.15$
& 0.48\%
& $\mathbf{0.73}$
& 0.32
& 36.1/54.2$^{*}$ \\

\midrule

\multirow{2}{*}{\textbf{Base}}
& $\mathcal{M}_{\mathrm{TTS}}$
& 5.91\%
& 0.33
& $\sim$
& $\sim$
& $\sim$
& $\sim$
& $\sim$ \\

& $\mathcal{M}_{\mathrm{TTS\text{-}DP}}$
& 6.48\%
& \underline{0.34}
& $\sim$
& $\sim$
& $\sim$
& $\sim$
& $\sim$ \\

\midrule

\multirow{3}{*}{\textbf{Ablations}}
& $\mathcal{M}_{\mathrm{TTS\text{-}SFT}}$
& 4.01\%
& 0.32
& $-1.08 \pm 0.20$
& 3.31\%
& 0.43
& 0.70
& 32.5/35.5 \\

& $\mathcal{M}_{\mathrm{TTS\text{-}SFT}}$ ($s=0.9$)
& \underline{3.21\%}
& 0.32
& $-0.98 \pm 0.19$
& \underline{3.22\%}
& 0.43
& 0.71
& 31.7/41.7$^{*}$ \\

& $\mathcal{M}_{\mathrm{TTS\text{-}DP\text{-}SFT}}$
& 4.82\%
& 0.32
& \underline{$-0.94 \pm 0.20$}
& 4.52\%
& \underline{0.44}
& 0.67
& 30.0/43.7$^{*}$ \\

\midrule

\multicolumn{2}{c}{\textbf{EmphTTS}}
& 3.30\%
& 0.32
& $-0.95 \pm 0.21$
& 3.25\%
& 0.42
& \underline{\textbf{0.75}}
& $\sim$ \\

\bottomrule
\end{tabular*}
\vspace{-3mm}
\end{table*}

\subsection{Baselines}

We choose four state-of-the-art open-source TTS systems as baselines: CosyVoice3~\cite{du2025cosyvoice}, FishAudio-S2~\cite{liao2026fish}, Qwen3-TTS~\cite{hu2026qwen3}, and VoxCPM2~\cite{voxcpm2_2026}, all of which support prosody control via explicit markup~\cite{du2025cosyvoice,liao2026fish} or voice design instructions~\cite{hu2026qwen3,voxcpm2_2026}. For CosyVoice3, we 
wrap emphasized words with \textit{\textless strong\textgreater}-\textit{\textless /strong\textgreater} tags as instructed. For FishAudio-S2, we use the \textit{[emphasize]} tag to indicate emphasized words. For Qwen3-TTS, we use the \textit{Qwen3-TTS-Voice-Design} (Qwen3-TTS-VD) model and experiment with different voice design instructions and text formats, finding that wrapping emphasized words with asterisks works best. We adopt the same markup for VoxCPM2 as the {Qwen3-TTS-VD} for the same reason. 

\subsection{Pre- \& Post-training}

$\mathcal{M}_\text{TTS}$ follows the \textit{F5TTS\_v1\_Base}\footnote{\url{https://github.com/SWivid/F5-TTS/blob/main/src/f5_tts/configs/F5TTS_v1_Base.yaml}} configuration and is pre-trained with a character-based vocabulary on an NVIDIA H200 GPU for 100k updates on the LibriTTS-R~\cite{koizumi2023libritts} dataset, with a batch size of 38400 mel-spectrogram frames. The duration predictor $\mathcal{M}_\text{DP}$ discretizes predicted durations into 100\,ms bins up to 30 seconds and shares the same input vocabulary as $\mathcal{M}_\text{TTS}$. It is pre-trained on the English subset of Emilia~\cite{he2024emilia} to handle diverse speaking styles, on 4 NVIDIA V100-32G GPUs with 48 utterances per GPU per batch.

At the SFT stage, we use the Expresso~\cite{nguyen2023expresso} dataset, in which emphasized words are labeled with asterisks. Since asterisks are absent from the LibriTTS-R vocabulary, we extend the text embeddings of both pre-trained models with a randomly initialized embedding for the asterisks. To prevent catastrophic forgetting, we mix a subset of pre-training data with Expresso: the \textit{train-clean-100} subset of LibriTTS-R for $\mathcal{M}_\text{TTS}$, and 17k randomly selected utterances from Emilia for $\mathcal{M}_\text{DP}$. $\mathcal{M}_\text{TTS}$ is fine-tuned for 100k updates with 38400 frames per batch, using a linear warmup over 20k steps to a peak learning rate of $1e^{-5}$, followed by linear decay, yielding $\mathcal{M}_\text{TTS-SFT}$. Similarly, $\mathcal{M}_\text{DP}$ is fine-tuned for 10k updates with 48 utterances per batch, using a linear warmup over 2000 steps to the same peak learning rate, followed by linear decay, yielding $\mathcal{M}_\text{DP-SFT}$.

For the GRPO stage, $\mathcal{M}_\text{TTS-SFT}$ is inferred with 16 sampling steps and a classifier-free guidance scale of 2.0. $\mathcal{M}_\text{DP-SFT}$ is optimized with a constant learning rate of $5e^{-6}$, batch size 1, group size $G=16$, and gradient accumulation of 8 steps, yielding the final $\mathcal{M}_\text{DP-GRPO}$. The reward weights are $\lambda_\text{WER}=3.0$, $\lambda_\text{SIM}=1.0$, and $\lambda_\text{BalanAcc}=1.0$ for Eq.~\eqref{eq:reward}. Both WER and BalanAcc are derived from WhiStress~\cite{yosha2025whistress}, a fine-tuned English-only Whisper-small model that jointly produces a transcription and word-level emphasis predictions. SIM is computed between the speaker embeddings of $\hat{\mathbf{y}}_i$ and $\mathbf{y}_\text{ref}$, extracted using CAM\texttt{++}~\cite{wang2023cam++}. The clipping value $\epsilon=0.2$ in Eq.~\eqref{eq:grpo_clip} and the KL divergence weight is $\beta=0.04$ in Eq.~\eqref{eq:grpo}. At this stage, we mix a randomly sampled 13k utterances from VCTK~\cite{yamagishi2019cstr} with Expresso, as pilot experiments showed this yields better emphasis generation than mixing \textit{train-clean-100} from LibriTTS-R with Expresso.

\subsection{Objective Evaluation}

We evaluate all systems on the Seed-TTS Eval~\cite{anastassiou2024seed} English subset for general-purpose TTS capability. The Seed-TTS Eval English subset contains 1088 prompt-target pairs for evaluating zero-shot TTS capability. The WER measures intelligibility between groundtruth transcripts and transcripts from \textit{Whisper-large-v3}~\cite{radford2023robust}. SIM is measured between the speaker embeddings of the prompt and synthetic speech, extracted using a WavLM-large-based~\cite{9747814} speaker verification model.

For emphasis controllability evaluation, we use the test split of TinyStress-15k~\cite{yosha2025whistress}, which includes 1000 synthetic utterances derived from TinyStories~\cite{Eldan2023TinyStoriesHS} with 10 voices and validated by humans. We keep 903 utterances with fewer than 3 emphasized words to match Expresso's distribution. One random utterance per speaker is selected from the TinyStress-15k training set as a prompt for TTS inference. The only exception is Qwen3-TTS-VD, which does not support voice cloning, and we allow random voice sampling in this case. The same objective evaluators are used as in Seed-TTS Eval. We additionally measure word-level emphasis using the F1 score from \textbf{StressLM}~\cite{yosha2025stresstest}: a Qwen2-Audio-7B~\cite{Qwen2-Audio} speech language model fine-tuned for emphasis detection and reasoning, and uses LLM-as-judge for prediction postprocessing. We run it in emphasis detection mode, which directly produces word-level predictions.




\subsection{Subjective Evaluation}
Two types of subjective evaluation are conducted: Comparative Mean Opinion Score (CMOS) and Emphasis Preference Test (EmphPref). For CMOS, we evaluate on Seed-TTS Eval. We randomly sampled 300 groundtruth utterances and paired them with the corresponding synthetic speech from each system. Participants rate which of two samples is more human-like on an integer scale from $-3$ to $3$, where $-3$ indicates \textit{Sample A is definitely more human-like}, $0$ indicates \textit{both are equally human-like}, and $3$ indicates \textit{Sample B is definitely more human-like}. One sample is the groundtruth and the other is synthetic. For EmphPref, we conduct eight comparisons in a one-versus-rest manner, with \textbf{EmphTTS} compared against groundtruth, each baseline, and each ablation. To reduce the confounding of intelligibility on emphasis preference, we only retain utterances for which the two compared systems have the same WER. Utterances are then randomly sampled for each comparison. After listening to both samples, participants select the one with better emphasis realization or indicate no preference.
For both evaluations, attention checks are embedded in the test, and submissions failing the attention checks are excluded from the statistics. We collect 25 valid submissions per evaluation from \textit{Prolific}\footnote{\url{https://www.prolific.com/}}.


\section{Results}

Experiment results are listed in Table~\ref{tab:exp}.

\textbf{Seed-TTS EN.}
Benefiting from large-scale training data and LLM backbones, all four baselines perform strongly on Seed-TTS EN, achieving WERs between 0.99\% and 1.84\%. {VoxCPM2} obtains the highest SIM of 0.753, while {Qwen3-TTS-VD} achieves the best CMOS of $0.06\pm0.17$. Overall, the baselines substantially outperform our smaller models in both intelligibility and speaker similarity.


Among our models, $\mathcal{M}_\text{TTS}$ achieves a WER of 5.91\% and a SIM of 0.33, while adding a duration predictor ($\mathcal{M}_\text{TTS-DP}$) further degrades WER to 6.48\%, suggesting a mismatch between the two components. SFT improves WER to 4.01\%, and the $s=0.9$ variant further reduces it to 3.21\%. $\mathcal{M}_\text{TTS-DP-SFT}$ obtains a WER of 4.82\%, whereas \textbf{EmphTTS} reduces it to 3.30\%, a 31\% relative improvement, indicating that GRPO effectively alleviates the mismatch introduced by duration conditioning. Subjectively, $\mathcal{M}_\text{TTS-DP-SFT}$ and \textbf{EmphTTS} achieve the best CMOS among our variants at $-0.94\pm0.20$ and $-0.95\pm0.21$, respectively, outperforming $\mathcal{M}_\text{TTS-SFT}$ and its $s=0.9$ variant.

\textbf{TinyStress-15k.}
On TinyStress-15k, the zero-shot baselines achieve substantially lower WERs and higher speaker similarities than our models. FishAudio-S2 obtains the lowest WER of 0.25\%, while VoxCPM2 achieves the highest SIM of 0.73. In comparison, our models obtain WERs between 3.22\% and 4.52\% and SIMs between 0.42 and 0.44. \textbf{EmphTTS} achieves a WER of 3.25\%, improving over both $\mathcal{M}_\text{TTS-SFT}$ (3.31\%) and $\mathcal{M}_\text{TTS-DP-SFT}$ (4.52\%), and is only marginally behind the best WER of 3.22\% obtained by the $s=0.9$ variant.

A different trend appears for emphasis controllability. The strongest zero-shot baseline, Qwen3-TTS-VD, reaches a StressLM F1 of 0.63, whereas all of our SFT-based models achieve higher scores. $\mathcal{M}_\text{TTS-SFT}$ reaches 0.70, and globally reducing the speaking rate with $s=0.9$ provides only a small improvement to 0.71. In contrast, \textbf{EmphTTS} achieves the best StressLM F1 of 0.75, outperforming the strongest baseline by 0.12 absolute F1. Together with its competitive WER, this indicates that GRPO improves emphasis realization without sacrificing intelligibility and that the gain cannot be explained by global speaking-rate reduction alone.

For the subjective EmphPref evaluation, Table~\ref{tab:exp} reports loss/win (L/W) percentages from the perspective of \textbf{EmphTTS}, where a win denotes a preference for \textbf{EmphTTS}. Statistical significance is determined using a one-sided binomial test on non-tie responses. \textbf{EmphTTS} is significantly preferred over groundtruth
(53.7\% vs.\ 36.0\%) and over CosyVoice3, FishAudio-S2, and VoxCPM2, with win rates of 56.0\%, 56.5\%, and 54.2\%, respectively. Its comparison with Qwen3-TTS-VD is near parity (45.7\% vs.\ 44.8\%), despite the substantially higher objective emphasis score of \textbf{EmphTTS}. The near-parity result with Qwen3-TTS-VD, despite the substantially higher objective emphasis score of \textbf{EmphTTS}, suggests that listeners' choices are also affected by the overall quality of the generated speech. This suggests considering objective metrics and subjective preference together when evaluating emphasis control. Among our ablations, \textbf{EmphTTS} is significantly preferred over $\mathcal{M}_\text{TTS-DP-SFT}$ (43.7\% vs.\ 30.0\%) and $\mathcal{M}_\text{TTS-SFT}$ ($s=0.9$) (41.7\% vs.\ 31.7\%), while the difference from $\mathcal{M}_\text{TTS-SFT}$ is not significant (35.5\% vs.\ 32.5\%). Taken together, the objective and subjective results show that GRPO improves the coordination between duration prediction and the frozen TTS model, learning input-dependent duration adjustments for emphasis rather than merely increasing the utterance duration globally.

\section{Conclusion}

In this paper, we present \textbf{EmphTTS}, an NAR TTS system for controllable word-level emphasis generation. We show that while SFT provides emphasis controllability, simply combining it with an independently trained duration predictor can be suboptimal at inference due to the mismatch between the two components. To address this mismatch, we optimize the duration predictor with GRPO using an emphasis reward based on balanced accuracy. \textbf{EmphTTS} achieves the best objective emphasis controllability and is significantly preferred over synthetic groundtruth samples and most baseline systems in subjective evaluation. Ablations further show that the gains from GRPO cannot be explained by a simple global reduction in speaking rate, supporting the importance of input-dependent duration optimization. We also observe that subjective emphasis preference can be influenced by overall speech quality, motivating the joint use of objective and subjective measures. Overall, our results demonstrate the potential of RL-based duration optimization for fine-grained prosodic control in TTS beyond emphasis.

\section{Acknowledgment}
This work was a part of the Ministry of Education and Culture’s Doctoral Education Pilot in Finland, under Decision No. VN/3137/2024-OKM-6 (The Finnish Doctoral Program Network in Artificial Intelligence, AI-DOC), and funded by the Strategic Research Council of Finland through the project ``Designing Inclusive \& Trustworthy Digital Public Services for Migrants in Finland (Trust-M)'' (Grant No.~353529). We acknowledge the computational resources provided by the Aalto Science-IT project.


\bibliographystyle{IEEEbib}
\bibliography{strings,refs}

\end{document}